# Antiphase Fermi-surface modulations accompanying displacement excitation in a parent compound of iron-based superconductors

Kozo Okazaki[1,*], Hakuto Suzuki[2], Takeshi Suzuki[1], Takashi Yamamoto[1], Takashi Someya[1], Yu Ogawa[1], Masaru Okada[1], Masami Fujisawa[1], Teruto Kanai[1], Nobuhisa Ishii[1], Jiro Itatani[1], Masamichi Nakajima[3], Hiroshi Eisaki[3], Atsushi Fujimori[2], and Shik Shin[1,*]
[1] *Institute for Solid State Physics (ISSP), University of Tokyo, Kashiwa, Chiba 277-8581, Japan*
[2] *Department of Physics, University of Tokyo, Bunkyo-ku, Tokyo 113-0033, Japan*
[3] *National Institute of Advanced Industrial Science and Technology (AIST), Tsukuba, Ibaraki 305-8568, Japan*
[*] *To whom correspondence should be addressed. E-mail: okazaki@issp.u-tokyo.ac.jp; shin@issp.u-tokyo.ac.jp*

We investigate the transient electronic structure of $BaFe_2As_2$, a parent compound of iron-based superconductors, by time- and angle-resolved photoemission spectroscopy. In order to probe the entire Brillouin zone, we utilize extreme ultraviolet photons and observe photoemission intensity oscillation with the frequency of the $A_{1g}$ phonon which is antiphase between the zone-centered hole Fermi surfaces (FSs) and zone-cornered electron FSs. We attribute the antiphase behavior to the warping in one of the zone-centered hole FSs accompanying the displacement of the pnictogen height, and find that this displacement is the same direction as that induced by substitution of P for As, where superconductivity is induced by a structural modification without carrier doping in this system.

Photo-irradiation to a material can trigger a phase change into crystallographically and/or electronically modulated structures. This phenomenon is known as photo-induced phase transition (PIPT) [1]. The observation of extremely efficient photochromism (bi-directional switching of molecular structures induced by photo-irradiation) in polydiacetylenes is a pioneering work of the discovery of PIPT [2], and emergence of the photo-induced 'hidden' insulating phase in manganese oxide [3] and photo-induced insulator-metal transition in vanadium dioxide [4] are known as typical examples of PIPT in strongly correlated materials.

Recently, it has been reported that photo-induced superconductivity is realized in cuprate superconductors and doped fullerene [5–7]. For 1/8-doped (La,Eu,Sr)$CuO_4$ (Eu-LSCO), where superconductivity is not observed in the equilibrium state, the Josephson plasma edge was observed by an optical pumping of mid-infrared pulse. This has been claimed as evidence for photo-induced superconductivity [5]. Another example is $YBa_2Cu_3O_{6+x}$ (YBCO), which has two $CuO_2$ layers within a unit cell and in which, differently from LSCO, two Josephson plasma edges and the transverse Josephson plasma mode due to the intra-bilayer and inter-bilayer couplings are observed. Surprisingly, the signal of the Josephson plasma has been detected even at room temperature [6, 8].

A key mechanism for the photo-induced superconductivity has been ascribed to atomic displacements induced by anharmonic effects of strongly excited infrared active phonon modes. Lattice modulation corresponding to the $A_g$ phonon has been confirmed by time-resolved X-ray diffraction (TRXRD) measurements using a free electron laser for YBCO [9, 10]. On the other hand, the observation of coherent phonon excitation indicates the existence of lattice modulations induced by optical pumping, since coherent phonon modes are excited followed by the lattice modulations according to the displacive excitation of coherent phonons (DECP) mechanism [11]. It has been reported that the photo-induced superconductivity can also be realized with near-infrared pump of 1.5 eV rather than mid-infrared pump in $(La,Ba)_2CuO_4$ [12]. This may be related to the lattice modulations through the DECP mechanism.

One of the most prominent features of $BaFe_2As_2$ (Ba122), the parent compound of another high-$T_c$ superconductor system, iron-based superconductors [13], is the emergence of superconductivity under various conditions; that is, hole doping by substitution of K for Ba [14], electron doping by substitutions of Co for Fe [15], and isovalent substitution of P for As [16], as well as under high pressure [17]. In the isovalent substitution and high-pressure studies, superconductivity can be induced by structural modifications without carrier doping. This leads to the idea that this material could be driven into a superconducting state by photo-irradiation. To this end, it is crucial to study the transient electronic structure after photo-irradiation.

In this Rapid Communication, we investigate the transient electronic structure of Ba122 by time- and angle-resolved photoemission spectroscopy (TARPES) to explore the possibility of photo-induced superconductivity in this system. An extreme ultraviolet laser from high harmonic generation is used to study both the hole and electron FSs, and we observe photoemission intensity oscillation after optical pumping, of which the frequency corresponds to the $A_{1g}$ phonon. We also find that the intensity oscillation is antiphase between the hole and electron FSs. We attribute this antiphase behavior to the warping [18] in one of the hole FSs accompanying the displacement of the pnictogen height. Since the pnictogen height is known to be important for superconductivity in iron-based superconductors [19] and the Ba122-based system shows superconductivity by the reduced pnictogen height [16], our observation should be important for the realization of photo-induced superconductivity in iron-based superconductors.

TARPES measurements were performed with a commercial extremely stable Ti:Sapphire regenerative amplifier system (Astrella, Coherent) with the center wavelength of 800 nm and pulse duration $\sim$ 30 fs, which was used for the pump light, and a Scienta R4000 hemispherical electron analyzer. After taking a second harmonic (SH) via 0.2-mm-thick $\beta$-$BaB_2O_4$ (BBO), the SH light is focused to the static gas cell filled with Ar and the high harmonics are generated. We selected 9th harmonic of the SH ($h\nu = 27.9$ eV) for the probe light using a set of SiC/Mg multilayer mirrors [20]. The probe light was $p$-polarized. (A component vertical to the surface was contained in the electric field.) While the other experimental conditions were similar to our previous report [21], we succeeded in observing a clear intensity oscillation corresponding to the coherent phonon excitation because the pulse duration of the fundamental laser was shortened and the stability was improved. The temporal resolution was evaluated to be $\sim$ 80 fs from the TARPES intensity far above the Fermi level corresponding to the cross correlation between pump and probe pulses. The total energy resolution was set to $\sim$ 250 meV. All the spectra were taken at 10 K. High quality single crystals of $BaFe_2As_2$ were grown by self-flux method and clean surfaces were obtained by cleaving *in situ*. Band structure calculations based on density functional theory (DFT) were performed using a WIEN2k package [22]. $a = 3.9625$ Å, $c$ = 13.0168 Å, and $h$ = 1.3602 Å [14] were used for the lattice parameters in equilibrium.

Figure 1(a) shows FS mapping measured with a He discharge lamp around the center and corner of the two-dimensional Brillouin zone (BZ), where the hole Fermi surfaces and electron FSs exist, respectively. The calculated FSs based on DFT are also overlaid. Figures 1(b) and 1(c) show the momentum integrated TARPES intensity measured across the hole and electron FSs, respectively, as a function of energy relative to the Fermi level ($E_F$) and pump-probe delay time. The integrated momentum regions of the hole and electron FSs are indicated by thick red and blue lines in Fig. 1(a), respectively. The spectra show that electrons are excited to above $E_F$ upon photo-irradiation and relaxed after that. In addition, oscillatory behaviors can be recognized for the intensity above $E_F$ in the spectra. This can be more clearly recognized in Figs. 1(d) and 1(e), where the integrated intensities of the regions surrounded by boxes in Figs. 1(b) and 1(c), respectively, are plotted. The dashed lines indicate the relaxation curves composed of an exponential decay function plus a residual slowly decaying component convoluted with a Gaussian corresponding to the temporal resolution. Now, the oscillatory components are clearly seen as being superimposed onto the background relaxation curve.

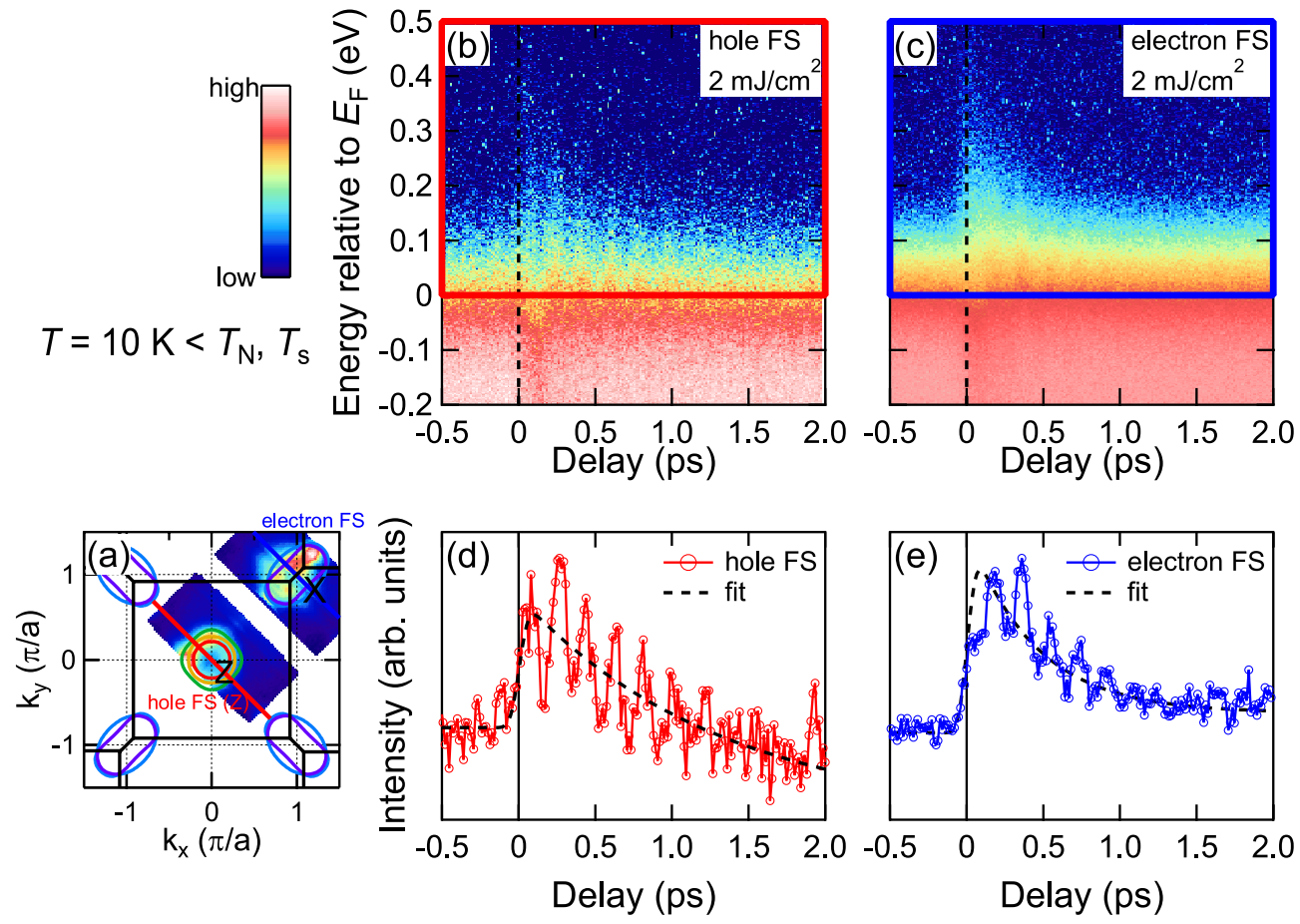

FIG. 1: Time-resolved ARPES (TARPES) spectra of $BaFe_2As_2$. (a) FS mapping measured with a He discharge lamp. The integrated momentum regions for the TARPES spectra across the hole and electron FSs shown in (b) and (c) are indicated by thick red and blue lines, respectively. The calculated FSs based on DFT are also overlaid. (b), (c) TARPES intensity images taken with the pump fluence of 2 mJ/cm$^2$ with respect to energy relative to $E_F$ and pump-probe delay (b) around the zone center and (c) around the zone corner in the two-dimensional BZ, respectively. (d), (e) Integrated intensity from $E_F$ to 0.5 eV above $E_F$ of spectra shown in (b) and (c), respectively. $T$, $T_N$, and $T_s$ are the measurement, Néel, and structural transition temperatures, respectively ($T_N \approx T_s$ = 142 K).

To gain further insight into the observed oscillatory behaviors, we show in Fig. 2(a) the differential curves between the integrated intensities and relaxation curves indicated by the dashed lines in Figs. 1(d) and 1(e), and their fast Fourier transform (FFT) in Fig. 2(b). The peak positions are located at $\sim$ 5.5 THz both for the hole and electron FSs, which evidences that the electronic system oscillates collectively with the frequency of the $A_{1g}$ phonon, as reported by the previous studies [23–26]. By comparing the top and bottom positions of the oscillatory components between the hole and electron FSs, it is noticed that their oscillatory behaviors are antiphase contrary to the previous report [27–29]. In order to confirm this, we have fitted the oscillatory components to the damped oscillation function;

$$(Ae^{-(t-t_0)/\tau} + B)\cos\omega(t-t_0),$$

where $t$ is the pump-probe delay time and $\omega$ is the frequency of the $A_{1g}$ phonon (= 5.5 THz), and the others are fitting parameters. The fitting results are shown in Figs. 2(c) and 2(d) for the hole and electron FSs, respectively, and the obtained values are $A$ = -0.22, $B$ = 0, $\tau$ = 1.3 ps, and $t_0$ = -10 fs for the hole FSs and $A$ = 0.32, $B = 0.02$, $\tau = 0.4$ ps, and $t_0 = 0$ fs for the electron FSs. From the facts that the sign of $A$ is opposite between the hole and electron FSs and $t_0$ is much smaller than the temporal resolution, we concluded that the phase of the oscillation is fully inverted with respect to each other.

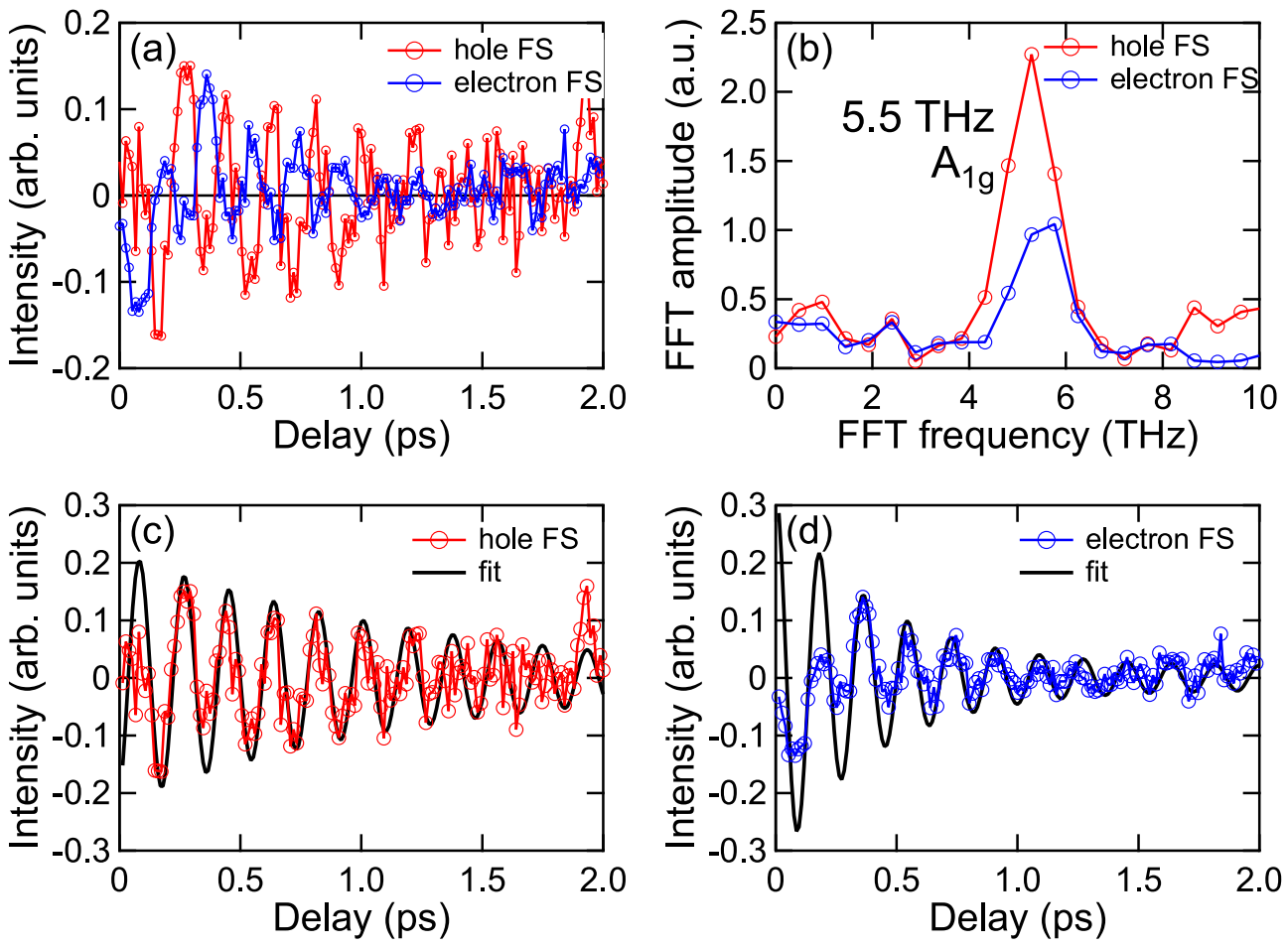

FIG. 2: Analysis of the oscillatory components of the TARPES spectra. (a) Oscillatory components of the hole and electron FSs deduced from the difference between the integrated intensities and decay functions shown in Figs. 1(d) and 1(e), respectively. (b) FFT amplitudes of the oscillatory components of the hole and electron FSs. (c), (d) Fitting results for the oscillatory components to the damped oscillator. Note that the phase of the oscillation is fully inverted between the hole and electron FSs.

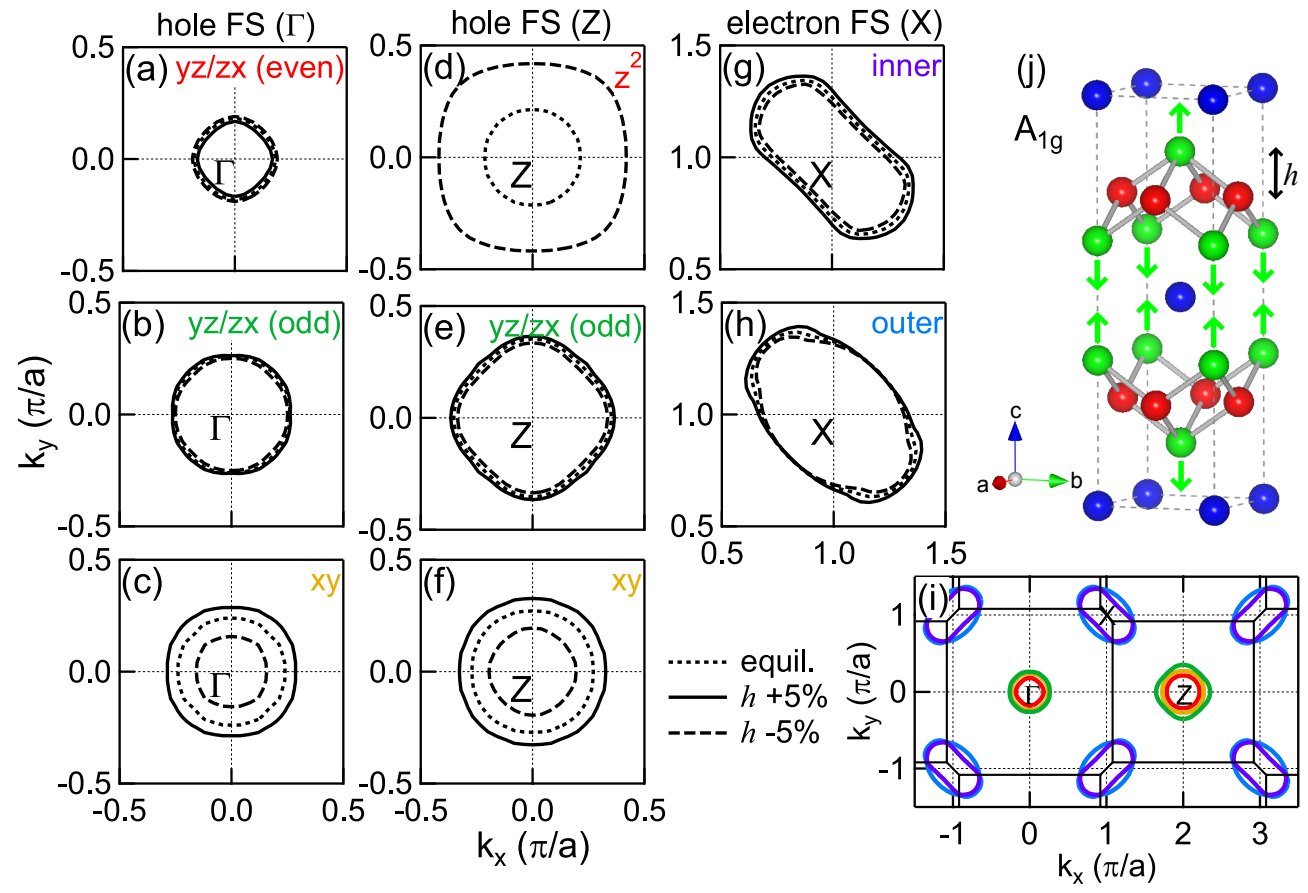

FIG. 3: Modulation of the electronic structure caused by the atomic displacement corresponding to the $A_{1g}$ phonon. (a)-(c), (d)-(f) Modulations of the hole FSs around the Γ and $Z$ points, respectively. (g), (h) Modulations of the electron FSs around the $X$ point. (i) Calculated FSs of the equilibrium state shown in the extended two-dimensional BZ including the Γ, $Z$, and $X$ points. (j) Crystal structure of $BaFe_2As_2$ and the definition of the pnictogen height $h$. Thick arrows indicate the displacements of the As atoms corresponding to the $A_{1g}$ phonon. Note that the $d_{xy}$ FS have low intensity due to matrix-element effects and will not be clearly visible by ARPES [30].

In order to investigate the origin of the phase inversion, we have performed band-structure calculations based on DFT for structures modulated by the $A_{1g}$ phonon [31]. Because the $A_{1g}$ phonon mode of Ba122 is the antiphase vibration of As atoms along the $c$ axis, one can simulate the electronic structure modulated by the $A_{1g}$ phonon mode with the modulation of the internal coordinate of the As atom $z$. We performed calculations for three values of $z$, where the pnictogen height $h$ defined as the distance of As atoms from the nearest Fe layer is in equilibrium and varied by $\pm 5$ % from the equilibrium value. The calculated hole FSs around the Γ and $Z$ points are shown in Figs. 3(a)-3(c) and 3(d)-3(f), respectively, and the electron FSs around the $X$ point are shown in Figs. 3(g) and 3(h) [32]. As can be recognized from the band dispersions, the most strongly modulated FS is the hole FS around the $Z$ point contributed from the $d_{z^2}$ orbital; that is, the $d_{z^2}$ FS is strongly warped for the lower $h$ value whereas the warping is dramatically weakened for the higher $h$ value. As for the other two hole FSs, the modulation of the hole FS contributed from the $d_{yz/zx}$ orbital is relatively weak, while the $d_{xy}$ hole FS has a significant modulation. According to the calculation of the photoemission matrix element, the matrix element of the $d_{xy}$ orbital is smaller for the energy of probe light (= 27.9 eV) due to the small emission angle of photoelectrons and the $d_{z^2}$ orbital is more sensitive to $p$-polarization than other orbitals [30]. Thus, contribution from the $d_{z^2}$ orbital is dominant for the temporal evolution of the TARPES intensity observed in Fig. 1(b). On the other hand, the modulations of the electron FSs are inverted with respect to that of the $d_{z^2}$ hole FS, though their changes are slightly smaller; that is, they become larger for higher $h$ and smaller for lower $h$. Thus, the observed antiphase oscillations between the hole and electron FSs are attributed to the FS modulations induced by the displacement of As atoms accompanying the $A_{1g}$ coherent phonon excitation [33].

The mechanism of the coherent phonon excitation has been often discussed and the DECP mechanism is one of the most likely explanations [11]. In this mechanism, the adiabatic energy potential is modified due to photoexcitations and has the minimum with the finite atomic displacement corresponding to the $A_{1g}$ phonon. As a result, the $A_{1g}$ phonon is excited instantaneously and coherently. This mechanism has two important predictions that the oscillation of the coherent phonon shows $\cos(\omega t)$ dependence and only those modes with $A_g$ symmetry are excited [11]. All the above features are consistent with our observations as shown in Fig. 2.

It should be worthwhile discussing possible temporally emerging electronic states. Figure 4 schematically describes the observed coherent phonon excitation and cor-responding modulations of the $d_{z^2}$ hole FS based on the DECP mechanism. As shown in Figs. 2(c) and 2(d), the oscillatory component of the TARPES intensity increases for the hole FSs, whereas that of the electron FSs decreases just after the arrival of the pump light. Correspondingly, the size of hole and electron FSs should increase and decrease, respectively, which results from the lowering of the pnictogen height. According to the DECP mechanism, the adiabatic potential after photo-excitation is modified such that it has a minimum at a smaller $h$. Here, in order to show the variation of the FS size more clearly, we assume the range of the variation of the pnictogen height $\Delta h$ down to -5 %. This $\Delta h$ corresponds to the lowering of the pnictogen height by 6.8 pm and the variation of the Fe-As-Fe bond angle by 2.6°. This variation is too large compared to the observed amplitude of the lattice modulation by TRXRD measurements [25, 26]. However, as shown in Fig. 4(j), Ba122 shows superconductivity by lowering the pnictogen height with $\Delta h \sim$ -1.3 %, which corresponds to the lowering of the pnictogen height by 2.0 pm. Therefore, if the atomic displacements minimizing the adiabatic potential after photo-excitation could be larger than this value and the accompanying lattice modulations persisted after the photo-excited electrons were relaxed to a metastable state, the observed transient electronic states with coherent phonon excitation may enter a photo-induced superconducting state. This can be expected for semimetals like iron-based superconductors, because it has been known that the lifetime of photo-excited carriers in indirect-gap semiconductors like Si is far longer than the time scale of cooling dynamics [34, 35], whereas they are comparable in direct-gap semiconductors like GaAs [36]. Whether the photo-induced superconducting state can be realized in Ba122, it has to be confirmed by terahertz time domain spectroscopy or TARPES measurements with higher energy resolution in the future. Finally, we should note that the sign of the variation of the pnictogen height deduced here is opposite to that deduced from TRXRD measurements [25, 26]. This is probably because our measurements were performed at 10 K, far below the structural transition temperature ($T_s$), whereas the TRXRD measurements for the observation of the $A_{1g}$ coherent phonon excitation were performed above $T_s$ as schematically shown in Fig. 4(k) [37]. In addition, the difference of the probing depth between TRXRD and TARPES may also be a reason for this opposite tendency.

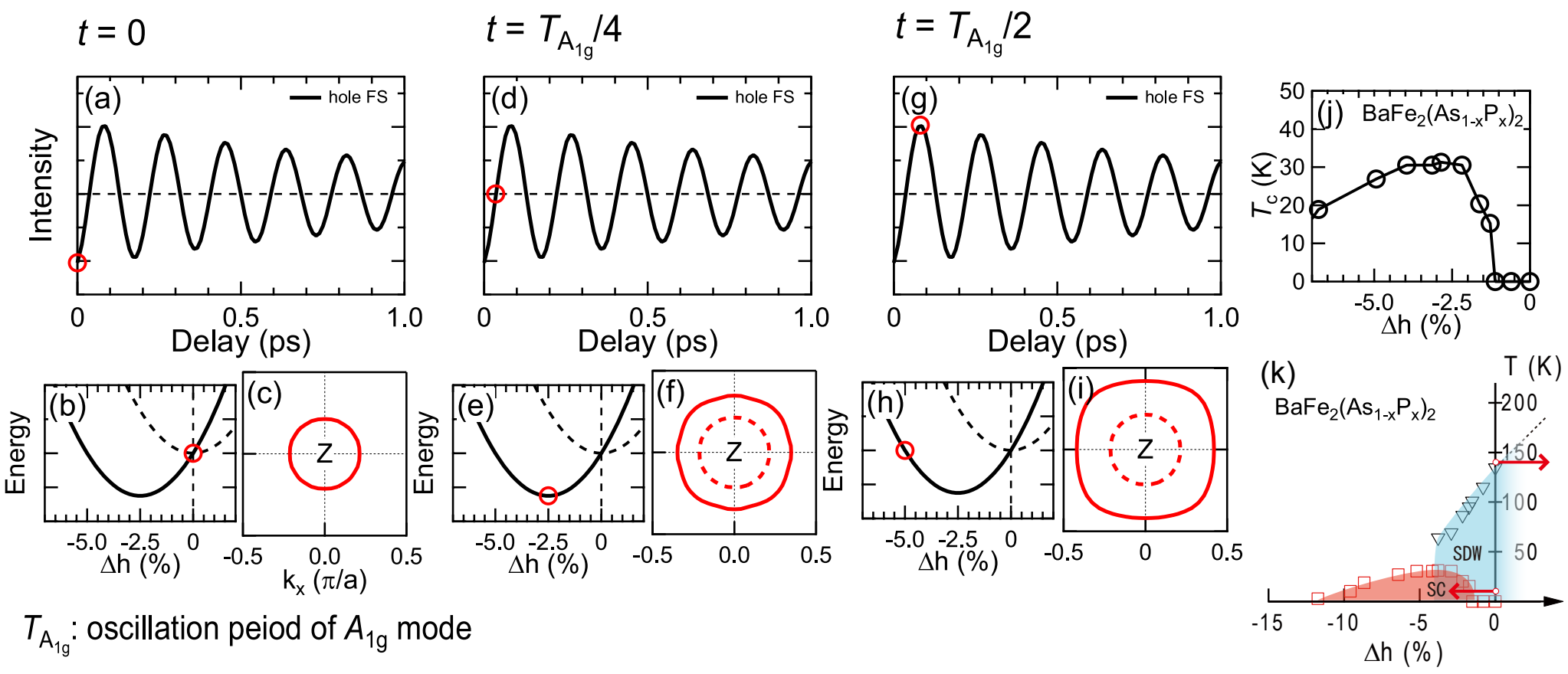

FIG. 4: Schematic description of the observed coherent phonon excitation and corresponding modulations of the $d_{z^2}$ hole FS. (a)-(c) Oscillatory component of the TARPES intensity for the hole FSs ($I_{\rm osc}$), the variation of the pnictogen height $\Delta h$, and the calculated $d_{z^2}$ FS, respectively, at $t = 0$. The dashed and solid lines in (b) indicate the adiabatic energy potential at the equilibrium state and the state after the coherent phonon excitation. At $t = 0$, $I_{\rm osc}$ is negative, $\Delta h = 0$, and thus $h$ is the largest and the size of the $d_{z^2}$ FS around the $Z$ point becomes the smallest. (d)-(f) The same as (a)-(c) but at $t = T_{A_{1g}}/4$, where $T_{A_{1g}}$ is the oscillation period of the $A_{1g}$ phonon. At $t = T_{A_{1g}}/4$, $I_{\rm osc}$ is neutral, $\Delta h$ is located where the energy is minimum, and the $d_{z^2}$ FS has a medium size. (g)-(i) The same as (a)-(c) but at $t = T_{A_{1g}}/2$. At $t = T_{A_{1g}}/2$, $I_{\rm osc}$ is positive, $\Delta h$ is negative and located where the energy is the same as that of $\Delta h$ =0, and the $d_{z^2}$ FS around the $Z$ point becomes the largest. The calculated $d_{z^2}$ FS at the equilibrium state is also shown as dashed lines in (f) and (i) for comparison. If this potential modulation that induces the coherent phonon excitation persists for some time, it is expected that the pnictogen atoms are relaxed to the negative $\Delta h$ positions and the warping of the $d_{z^2}$ FS remains enhanced. (j) $T_c$ of $BaFe_2(As_{1-x}P_x)_2$ plotted as a function of variation of the pnictogen height $\Delta h$. In $BaFe_2(As_{1-x}P_x)_2$, the pnictogen height lowers as P is substituted for As, and superconductivity emerges at $\Delta h \sim$ -1.3 % ($x \sim 0.16$). (k) Phase diagram of $BaFe_2(As_{1-x}P_x)_2$ with respect to $T$ and $\Delta h$. The signs of $\Delta h$ from TARPES and TRXRD are indicated by red arrows. The transition temperatures were taken from Ref. 16.

In conclusion, we have performed TARPES measurements on a parent compound of iron-based superconductors, $BaFe_2As_2$, using an extreme ultraviolet laser from high harmonic generation. The antiphase photoemission intensity oscillation between the hole and electron FSs was observed and attributed to the warping in the $d_{z^2}$ FS around the $Z$ point. We conclude that this displacement is the same direction as that induced by substitution of P for As, where superconductivity is induced by a structural modification without carrier doping. We also suggested a longer lifetime of the photo-excited state in semimetals like iron-based superconductors.

We would like to thank T. Tohyama, and M. Imada for valuable discussions and comments, and T. Yoshida for valuable comments. This work was supported by JSPS KAKENHI (Grant No. JP25220707 and JP26610095) and Photon and Quantum Basic Research Coordinated Development Program from the Ministry of Education, Culture, Sports, Science and Technology, Japan. H.S., T.S., and M.O. acknowledge the JSPS Research Fellowship for Young Scientists, and H.S. acknowledges Advanced Leading Graduate Course for Photon Science (ALPS) for financial supports.

# Supplemental Material

### Evidence for intensity oscillation rather than chemical potential oscillation

In the previous report, it has been reported that the effective chemical potential oscillates *in phase* between the Γ and $M$ points as a function of pump-probe delay time[S1], contrary to our observation of the antiphase oscillation. In this work, we have observed the oscillation of the photoemission intensity rather than the chemical potential. Figure S1 shows the temporal evolution of the integrated TARPES intensities on the electron FSs in different integration ranges and their oscillatory components. The stronger time-dependent modulation in our TARPES data than those of Ref. S1 allows us to perform a qualitative intensity analysis. Irrespective of the integration range, the phase of the oscillatory components does not change. This is strong evidence that the photoemission intensity itself oscillates rather than the chemical potential, because if the chemical potential oscillated, the whole spectra would be shifted accordingly and the oscillation for the photoemission intensity would be out of phase between the higher and lower energy side of a spectral peak. In the case of the iron-based superconductors, the peak position of the spectra is located just below $E_F$ due to the existence of the Fermi cutoff. If the chemical potential oscillated, the phase of the oscillatory component should be inverted between the integration ranges of [0.0, 0.5] eV and [-0.5, 0.0] eV.

### Band dispersions for the modulated structures

Figures S2(a) and S2(b) show the band dispersions calculated for three values of $h$, which are in equilibrium and varied by ±5 %. The modulation around the Z point is the most pronounced.

### Modulations of hole FSs at various $k_z$ values

While we attributed the antiphase behavior between the hole and electron FSs to the warping in the $d_{z^2}$ FS, which is expected to be the strongest at the Z point, the probe energy of 27.9 eV does not exactly trance the Z point in the Brillouin zone. According to the previous reports[S2,S3], $k_z$ traced by 27.9 eV photons is expected to be around the midpoint between the Γ and Z points. In order to assess this point to our conclusion, we calculated the FS modulations at various $k_z$ values as shown in Fig. S3. Based on these calculations, we could confirm that the modulation of the $d_{yz/zx+z2}$ hole FS (Whether the dominant contribution to this FS is $d_{yz/zx}$ or $d_{z2}$ orbital, it switches depending on $k_z$.) is antiphase with respect to those of the electron FSs for any $k_z$ value, while the modulation is the smallest at the Γ point and the largest at the Z point. In this regard, since $k_z$ traced by 22.1 eV used for Ref. S1 is very close to the Γ point, this could be a possible reason for the difference from the present results.

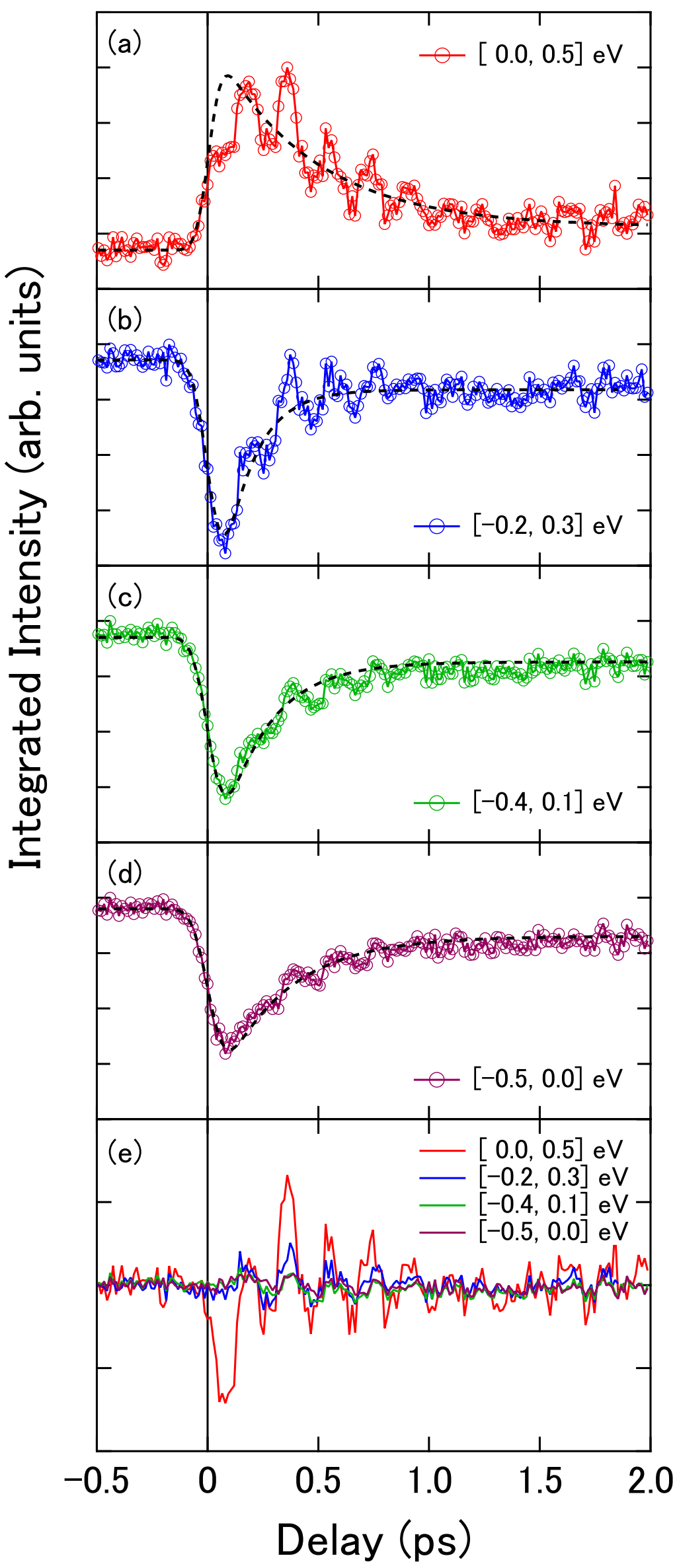

FIG. S1: Temporal evolution of the integrated photoemission intensities on the electron FS with various energy integration ranges and their oscillatory components. The integration ranges are (a) [0.0, 0.5] eV, (b) [-0.2, 0.3] eV, (c) [-0.4, 0.1] eV, and (d) [-0.5, 0.0] eV. The dashed lines indicate the decay function to extract the oscillatory components. (e) Deduced oscillatory components. Irrespective of the integration range, the phase of the oscillatory component does not change.

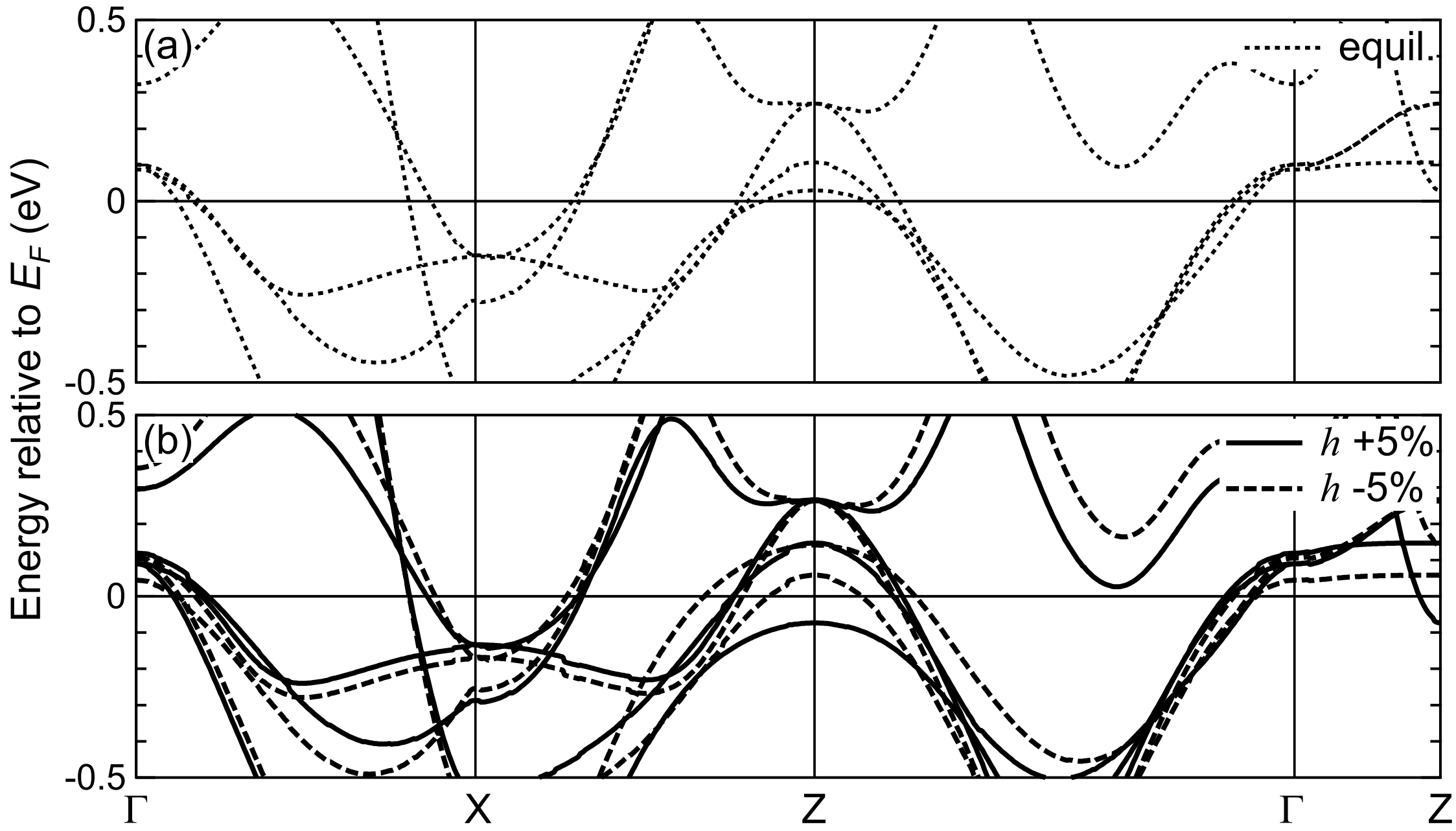

FIG. S2: Band dispersions of the modulated structures caused by the atomic displacement corresponding to the $A_{1g}$ phonon. (a) Band dispersions of the equilibrium state. (b) Those of the modulated structures.

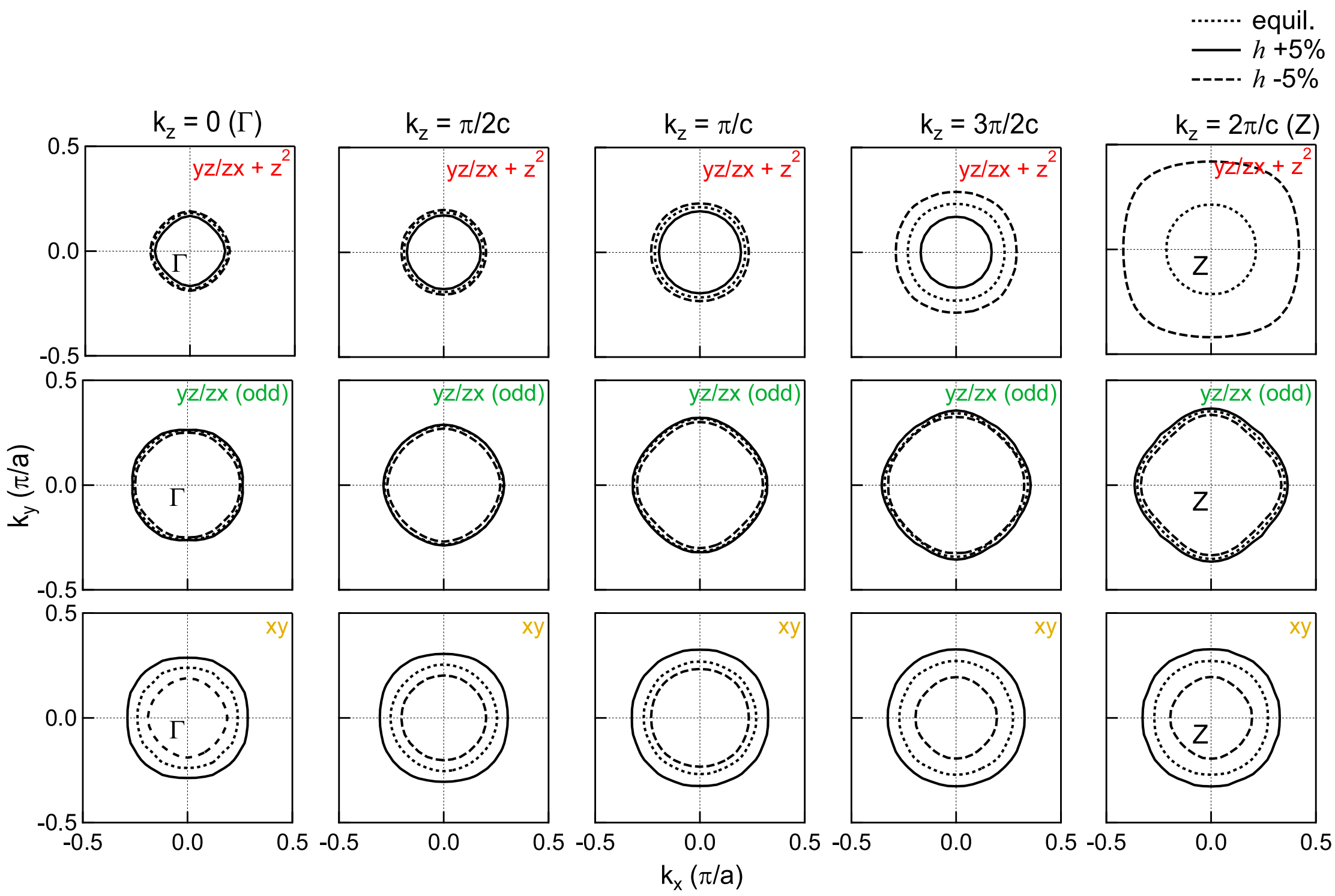

FIG. S3: Modulations of the hole FSs around the zone center at various $k_z$ values. Notations are the same as those of Fig. 3(a)-3(f).

### Animation of FS and structural modulations due to coherent phonon excitation

Supplemental Movie S1 includes an animation of FS and structural modulations due to the coherent excitation of $A_{1g}$ phonon based on the DECP mechanism.

Movie S1: Animation of FS and structural modulations. In order to show the modulations more clearly, we assume the range of the variation of the pnictogen height $\Delta h$ down to -10 %.